\documentstyle[12pt]{article}
\def \medio  {\baselineskip= 1.5 \normalbaselineskip}

\def \s  {\sigma}

\def \n  {\nu}
\newcommand{\titul}[1] {\begin{center}{\large\bf #1 } \end{center}\vskip 1.cm}
\newcommand{\autor}[1] {\begin {center} {\large \lineskip .5em #1 }
                        \end   {center} }
\newcommand{\lugar}[1] {\begin{center} {\it #1} \end{center}}
\newcommand{\abstr}[1] {{\begin{center} \vskip .5cm {\bf Abstract
                        \vspace{0pt}} \end{center}}\begin{quote} #1
                        \end{quote}}

\begin{document}
\begin{titlepage}

\begin{flushright} {\bf US-FT/9-00} \end{flushright}

\vskip 3.cm
\titul{
Measuring the BFKL Pomeron in Neutrino Telescopes
}
\autor{J.A. Castro Pena, G. Parente and E. Zas}
\lugar{Departamento de F\'\i sica de Part\'\i culas\\
Universidade de Santiago de Compostela\\
15706 Santiago de Compostela, Spain}
\abstr{
\medio
We present a new method for obtaining information on
the small $x$ behavior of the structure function $F_2$ outside
the kinematic range of present acelerators
from the mean inelasticity parameter in UHE neutrino-nucleon
DIS interactions which could be measured in neutrino telescopes.
}
\end{titlepage}
\newpage

\pagestyle{plain}
\medio
\section{Introduction} \indent

A variety of models have been proposed in which astrophysical neutrinos 
exceed the expected atmospheric neutrino flux at energies in the TeV 
scale or higher \cite{nufluxes}.  
The detectors presently in design or construction stages \cite{AMANDA}
look for \v Cerenkov light from the neutrino-induced muon in deep inelastic
scattering (DIS) charged 
current (CC) interactions and take advantage of both the long muon range 
and the rise of the neutrino cross section to meet the neutrino detection
challenge.

Several alternatives have been proposed for observing 
high energy neutrinos and they are all based on the detection of 
the showers that are also produced in most of the neutrino interactions. 
These include the detection of particle shower fronts in the atmosphere 
in the horizontal direction (Horizontal Air Showers) 
\cite{HAS} and coherent pulses from showers in different media, 
both acoustic \cite{radiosound} and in radio \cite{radiosound,avz99} waves.  
Shower detection is sensitive both to 
neutral current interactions and to all neutrino flavors.
In DIS interactions with nuclei, 
showers of hadronic type are always initiated by the debris of 
the struck nucleons. For electron neutrino charged current 
interactions an electromagnetic shower is initiated 
at the lepton vertex in addition to the hadronic shower. 
Certainly conventional detectors in 
construction can also look for the \v Cerenkov light from 
the induced showers provided they are contained or sufficiently 
close to the instrumented volume. In fact it is likely that 
shower detection is the main way these detectors can observe 
ultra high energy neutrinos because of the earth's opacity 
and the atmospheric muon background.

The high energy neutrino cross section is a crucial ingredient in the
calculation of the event rate in high energy neutrino telescopes. 
The DIS cross section is given in terms of structure functions
whose energy dependence (scaling violations) is calculated
in perturbative QCD. In practice, structure functions are computed
from available parton (quark and gluon) distribution functions
which depend on $x$ and $Q^2$ (see below).
Parton densities are extracted from data using DGLAP evolution
equations \cite{DGLAP} which effectively sum
$\ln(Q^2)$ terms contained in the QCD perturbative expansion.

The DGLAP approach can break down at low $x$ because 
of potentially large $\ln(1/x)$ terms which also
appear in the perturbative series.
The theoretical behaviour of structure functions at small $x$ 
has been discussed since the seventies and recently revived
in the context of the low $x$ HERA data 
(see Ref. \cite{CoDeRo} for a review).

The sumation of the $\ln(1/x)$ terms for intermediate $Q^2$
is performed with the BFKL formalism \cite{BFKL}
which evolves parton densities unintegrated in transverse momentum.
Within BFKL at LO and fixed $\alpha_s$, 
the structure function $F_2$ behaves as power like
$x^{- \lambda}$, with $\lambda \simeq 0.53$ for $\alpha_s$=0.2
(for $\alpha_s(M_z^2)$=0.12, $\lambda \simeq 0.32$).
The exponent $\lambda$ is related to the
intercept of the Pomeron which governs the high-energy asymptotics
of the total cross section.
The consideration of NLO
effects in BFKL is a matter of present discussion due to the large
correction found ($\lambda$ becames negative). However, recently it has
been found \cite{KIM} that the NLO value of $\lambda$, improved by
optimization of the renormalization
scale, has a very weak dependence on
the virtuality $Q^2$, ranging the values $\lambda \simeq 0.13-0.18$
at $Q^2=1-100$ GeV$^2$.

This result, $x^{- \lambda}$, is obtained in DGLAP when the input
structure function at fixed $Q_0^2$ has a singular behavior at low $x$
but also when it is flat. In the latter case, the solution
is the so call double logarithmic
approximation (DLA) and effectively resums terms of type
$\alpha_s \ln(1/x) \ln(Q^2)$ \cite{Rujula}. 
After evolution over a suficiently large $Q^2$ interval,
the DLA result mimics the BFKL behavior, $x^{- \lambda}$ \cite{NPB}. 

At this point it should be stressed that the low $x$ behavior of
structure functions
is fundamental in the
high energy limit of the neutrino-nucleon DIS cross section.
The calculation of the neutrino-nucleon cross section 
involves an integration over $x$ which 
corresponds to the fraction of momentum 
carried by the struck parton (see below).
For $x^{- \lambda}$ with $\lambda > 0$, 
as the neutrino energy increases, 
the integral becomes dominated by
the interaction with partons of lower $x$, 
while the $Q^2$ integral remains dominated by $Q^2$ values up
to the electroweak boson mass squared
(see for example \cite{ICRCDURBAN} and below). For $Q^2$ above
$M_W^2$ the integrand behaves as $Q^{-4}$ and quickly becomes
irrelevant.

On the experimental side there are however no $F_2$ 
measurements at very small $x$ and large $Q^2$. 
Consequently, in the process of the total cross section 
calculation, the parton densities extracted from present data
must be carefully extrapolated to the region in the 
$x, Q^2$ plane without experimenta support. 
The dominant kinematical region calls 
for the connection between the low $x$ behavior 
at intermediate $Q^2$ (the BFKL region) and the high 
$Q^2$ region but at moderate low $x$ obtained
in fits to experimental data (the DGLAP region) \cite{Martin2}. 

The consideration of QCD effects in
the neutrino DIS cross section for neutrino energies well above
the electroweak boson
threshold was done at the end of the seventies
\cite{AndrBerezSmir,Halprin}
motivated by the DUMAND project. Previously the neutrino-nucleon DIS
cross section had been studied in detail for lower neutrino energies
in an attempt to disentangle
asymptotic freedom effects and, in particular, scaling violations
(see \cite{Buras} for a review).
The study of the uncertainty in the cross section 
calculation, which is important in the context of 
high energy neutrino telescopes, has been recently 
addressed \cite{Gluck, Reno, Martin}. 
The uncertainty due to the extrapolation 
to high $Q^2$ is not expected to be large because the 
low $x$ shape of the parton distribution functions at low 
$Q^2$ is narrowly constrained by HERA while the further
evolution to high $Q^2$ values seems well defined by
DGLAP QCD evolution. 
At the highest energies uncertainties within
$20\%$ \cite{Gluck}, $40\%$ \cite{Martin} and a factor 
$2^{\pm 1}$ \cite{Reno} are typically reported. However we think
that the consideration of conventional DGLAP evolution to small
$x$ values and high $Q^2$, as for example
$x=10^{-9}$ and $Q^2=10^4$ GeV$^2$, should at least be questioned in view
of absence of data.

Conversely if any neutrino interactions measurements could be made 
leading to information on the cross section, 
such information would be of great value to constrain the
theoretical predictions at low $x$ and high $Q^2$.
Of particular interest is the $y$ distribution
which can be shown very sensitive to the value of $\lambda$ (see below),
$y$ being the fraction of 
the neutrino energy which flows to the hadronic part of the interaction 
in the laboratory frame.

This distribution is incidentally very important for neutrino detection,
particularly  for all the alternatives in neutrino detection which rely on
detecting the high energy showers which are always produced at the hadronic 
vertex, whatever the neutrino flavor and for both charged current (CC)
and neutral current (NC) interactions. 

The article is organized as follows:
Firstly we collect the most relevant formulae used in
the calculation of the neutrino nucleon cross section. 
Then, we present the result for the total cross section and the
average inelasticity $\langle y \rangle$ computed from two
sets of parton distribution functions. 
We explain the characteristics of the results for neutrinos, antineutrinos
in CC and NC interactions in terms of the parton distribution dependences,
stressing the connection between these quantities and the low $x$ limit
of the structure functions at the highest energies.

In the assumption that the low $x$ behavior of $F_2$ at large $Q^2$
can be described by $x^{-\lambda}$,
we present a very simple analytical
relation between ${\lambda}$ and the average
inelasticity $\langle y \rangle$ which does not
depend of any other parameter.
We suggest that ${\lambda}$ could be determined
from $y$-measurements in the events which are expected in high
energy neutrino detectors. 
Although $y$-measurements could be thought to be rather
speculative at this early stage neutrino astronomy is at present,
we point out a method to measure $y$ in neutrino telescopes.

\section{The neutrino-nucleon DIS cross section at UHE} 
\indent

In terms of structure functions the charged-current (CC) neutrino-nucleon
DIS differential cross section is given by:
\begin{eqnarray} 
\nonumber \\ 
\frac{d\sigma^{\nu (\overline{\nu})N}}{dx dy}  = 
  \left( \frac{G_F^2}{4\pi} \right) 2 M E_{\nu} 
  \left( \frac{M_W^2}{M_W^2 + Q^2} \right)^2 \; 
  ( y_{+} F_2 
 -  y^2 F_L \pm y_{-} x F_3 ) 
\label{eq:1}
\\ \nonumber
\end{eqnarray}
where $y_{\pm} = 1 \pm (1-y)^2$, $M$ is the nucleon mass,
$E_{\nu}$ the neutrino energy in the lab frame, 
$Q^2 = 2 M E_{\nu} x y $ and we have neglected terms suppressed by
powers of $M^2/Q^2$.

In the QCD improved parton model,
the structure functions $F_i$ ($i=2,3,L$) are calculated in terms
of quark and gluon distribution functions.
At leading order (LO) approximation in perturbative QCD $F_2$ ($xF_3$)
is simply related to the sum (difference) of parton densities.
At next-to-leading order (NLO) further integrals are involved relating
at order $\alpha_s(Q^2)$ the parton densities with the structure function.
The contribution due to $F_L$, which is also proportional
to $\alpha_s(Q^2)$ becomes small at large $Q^2$ and 
like the $M^2/Q^2$ terms are neglected in this work.

We present, for simplicity, the
expressions which relate the cross sections and the parton densities
at the LO approximation.
Within the above approximations the differential cross section
for the charged current (CC) interaction of neutrinos on isoscalar target
takes the form:
\begin{eqnarray}
\frac{d^2\s^{{\n}N}}{dxdy} & = & 
    \frac{G_F^2}{\pi} \frac{ M_W^4 M E_{\nu}}{(M_W^2 + Q^2)^2}
\left[ 
A(x, Q^2) + \overline{B}(x, Q^2)
(1-y)^2 
\right]
\nonumber \\
\frac{d^2\s^{{\bar \n}N}}{dxdy} & = &
    \frac{G_F^2}{\pi} \frac{ M_W^4 M E_{\nu}}{(M_W^2 + Q^2)^2}
\left[
\overline{A}(x, Q^2) + B(x, Q^2)
(1-y)^2 
\right]
\end{eqnarray}
where 
\begin{eqnarray}
A(x,Q^2)=x(u + d + 2s + 2b) \; , \;\;\;
B(x,Q^2)=x(u + d + 2c + 2t)
\nonumber
\end{eqnarray}
The antiquark combinations $\overline{A}$
and $\overline{B}$ are obtained from $A$ and $B$ replacing each quark
by the corresponding antiquark of the same flavor.

The total cross section is calculated by integration of Eq. (2).
We have used two representative sets of parton 
densities\footnote{The consideration of nuclear effects in parton
distributions which could be relevant for heavy nuclei will be presented
elsewhere \cite{CPZnuclear}}:
the LO set from Ref. \cite{MRST98} (MRST98) and the NLO set in the
DIS factorization scheme from Ref. \cite{GRV98} (GRV98)\footnote{ Rigurously,
NLO corrections modify Eq. (2) in the DIS scheme by terms proportional to
$\alpha_s$ coming from x$F_3$. We neglect them in this calculation}.
In the integration we extrapolate the quark
distribution functions in $x$ below the low $x$ limit given by the
authors ($x = 10^{-5}$ for MRST98 set and $x = 10^{-9}$ for the GRV98 set)
using the $x$ slope at the lowest
$x$ value of the parametrization for each value of $Q^2$.
This simple phenomenological extrapolation agrees
with the more elaborated prescriptions based on perturbative
QCD as for example the double-logarithmic-approximation which explain
HERA data (see \cite{ICRCDURBAN} and also Fig. 1b). 
We choose this ad hoc extrapolation for simplicity because
our purpose is just to show the sensitivity of the high energy
neutrino cross section to the low $x$ parton behavior.

The total charged and neutral current interaction
cross sections\footnote{For brevity, in this work we do not present
the explicit expressions used in the calculation of the NC cross sections}
for both neutrinos and anti-neutrinos calculated with MRST98 partons
are shown in Fig. 1a as a function of the neutrino energy.
At low energy one can observe the linear rise of the cross section
with the neutrino energy ($E_{\nu}$). At high energy it
would follow a logarithmic rise $\ln(E_{\nu})$ were it not for the
increase of the sea quark densities at small $x$ ($\sim x^{-\lambda}$)
which imply a high energy cross section that rises as
$(E_{\nu})^{\lambda} \; \ln(E_{\nu})$ for $E_{\nu}$ above 1 PeV.

We have also calculated the cross sections with GRV98 partons.
For neutrino energies considered in this work, i.e. below $10^{13}$ GeV,
the results are insensitive to the extrapolation below  $x=10^{-9}$.
The cross sections from MRST98 and GRV98 parton distribution
can be compared in Fig.1b.

\section{The mean inelasticity at UHE} 
\indent

In this work we are interested in the fraction of 
the neutrino energy which flows to the hadronic part of the interaction 
in the laboratory frame, $y$, which is also called inelasticity. 
For a given muon neutrino high energy flux, this parameter fixes
the relative rates of
the two main types of detections, using muons in charged current 
interactions and using the showers produced in the interactions.
It is also reponsible for the 
relative sizes of the electromagnetic and hadronic showers induced
in charged current electron neutrino interactions.
This quantity is necessary for any attempt at extracting the neutrino 
energy in high energy neutrino telescopes
whether by detecting the shower produced in the interaction
or by detecting the muon produced by muon neutrino CC interactions
 from the detected 
hadronic shower or muon.

The average value of $y$ can be obtained by integration of the 
differential cross section:
\begin{eqnarray}
\langle y \rangle = \frac{1}{\sigma} \int_0^1 dy \, y \, \frac{d\sigma}{dy}
\label{ymedio} 
\end{eqnarray}
The energy dependence of $\langle y \rangle$
has been studied in the past to test
asymptotic freedom in strong interactions (see \cite{Buras} for a review).
The effect of the boson propagator on
the $y$ distribution and the relevance of the small $x$ region were
pointed out in Refs. \cite{AndrBerezSmir,Halprin}.
 
In the present work we have calculated $\langle y \rangle$ using
modern parton densities which better describe the
low $x$ kinematic region. The results using MRST98 partons
are shown in Fig. 2a.

Let us explain the general features observed in Fig.~2a
from the cross section formulas given in Eq.~(2).
At low energy, the value of $\langle y \rangle$ for $\nu$
is larger than for $\bar{\nu}$ because the most important contribution
comes from the valence quarks which are suppressed by $(1-y)^2$ in the
anti-neutrino cross section formula (see Eq.~(2)).
If there were not sea quarks, the $y$ distributions given by Eq. (2)
would behave as $(1-y)^2$ (would be flat) for antineutrinos (neutrino) and
correspondingly $\langle y \rangle = 0.25$ ($0.5$).
In the case of low energy  $\bar{\nu}$, the increase of
$\langle y \rangle$ with energy is due to the rise of the sea quark
distribution functions. For neutrinos, this effect is less apparent. 

The depletion of $\langle y \rangle$ for
energies above 1000 GeV (see Fig. 2a)
is due to the $W$ propagator appearing in Eq. (2) which
acts as a cutoff in the integration restricting the values of $Q^2$ to
around $M_W^2$, i.e. $ x y \sim M_W^2/(2 M E_{\nu}) $. 
Thus, the $d\sigma/dy$ distribution is shifted to lower values of $y$.

At the highest energies, because of the sea quark dominance,
both neutrino and antineutrino cross sections become almost equal
(see Fig. 2a) and the
small difference between CC and NC interactions is due to the characteristic
couplings of the electroweak interaction.

Concerning the value of $\langle y \rangle$, Fig. 2a shows
that it typically becomes stable at values slightly above 0.2. 
It is remarkable that in the high energy limit
$\langle y \rangle$ should be constant while the average
value of the Bjorken $x$ variable, $\langle x \rangle$,
decreases strongly (see Fig. 4b).
The low $x$ rise of the sea
quark densities (predicted by perturbative QCD) shifts 
the weight of the cross section to small $x$ values compensating
the $y$ distribution.

The energy dependence of $\langle y \rangle$
at high energy is related to the behavior of the slope
assumed constant in the extrapolation of the MRST98 quarks at low $x$.
If $\lambda$ were negative or zero, $\langle y \rangle$ would
send to zero in this limit.

For example, if in the cross section calculation one uses 
GRV98 partons, $\langle y \rangle$ is not constant
and we find that it decreases with increasing
energy (see Fig. 2b) which reflects the depletion of the effective
slope $\lambda$ of the GRV98 partons with increasing $Q^2$.

In the following we show that the observed correlation between 
$\langle y \rangle$ at high energy and $\lambda$ at low $x$
can be put in analytic form using very simple approximations.

\section{Low $x$ physics and high energy neutrino telescopes}
\indent

Let us assume that $F_2$ at
low $x$ and high $Q^2$ is given by the power-like expression
$F_2=A(Q^2) \; x^{-\lambda}$, where $\lambda$ is assumed constant
(or smoothly dependent on $Q^2$ as predicted by \cite{KIM}) 
Then the CC neutrino-nucleon 
differential cross section (in $y$ and $Q^2$) takes the
form:
\begin{eqnarray}
\frac{d^2\sigma}{dy dQ^2}=
\left(\frac{G_F^2}{4\pi}\right)
\left(\frac{M_W^2}{M_W^2 + Q^2}\right)^2 A(Q^2)
\left(\frac{2ME}{Q^2}\right)^{\lambda}
\frac{[1+(1-y)^2]}{y} \; y^{\lambda}
\end{eqnarray}   
where in Eq. (1) we have neglected the contribution from
$F_L$ and $xF_3$ structure functions, expected to be negligable at
low $x$ and high $Q^2$.

In the calculation of $\langle y \rangle$ by Eq. (3) 
the $Q^2$ integral of Eq. (4) in numerator
and denominator cancels at high energy, i.e. provided that
$2ME_{\nu}$ is much larger than $M_W^2$. Then one has:
\begin{eqnarray}
\langle y \rangle = 
\frac{\int_0^1 dy \; y \; \frac{d\sigma}{dy}}{\int_0^1 dy \;
\frac{d\sigma}{dy}}  \simeq
\frac{\int_0^1 dy \; [1+(1-y)^2] \; y^{\lambda}}{\int_0^1 dy \; [1+(1-y)^2]
\; y^{\lambda-1}}
\end{eqnarray}
which can be easily integrated to get the simple analytical relation:
\begin{eqnarray}
\langle y \rangle \simeq 
\frac{\lambda^3 + 5\lambda^2 + 8\lambda}{\lambda^3 + 6\lambda^2 + 13\lambda
+ 12}
\end{eqnarray}

Analogously, one can proceed as above in the calculation of 
the average $Q^2$ which is expected (see above) to be around
$M_W^2$ at energies sufficiently above the $W$ propagator threshold.
One finds:
\begin{eqnarray}
\langle Q^2 \rangle = 
     \frac{1}{\sigma} \int dQ^2 \; Q^2 \; \frac{d\sigma}{dQ^2}
\simeq 
\frac{1-\lambda}{\lambda} M_W^2
\end{eqnarray}

To check Eq. (6), let us find out the value of $\lambda$ which
corresponds to $\langle y \rangle = 0.23$,
which is the result from the explicit integration of the CC
neutrino-nucleon differential cross section $d\sigma/dy$ calculated
with MRST98 parton densities at $E_{\nu}=10^{10}$ GeV (see Fig. 2a).
Looking at Fig. 3 one finds
$\lambda = 0.42$.
We have checked that the MRST98 sea partons at
$x=10^{-5}$ and $Q^2=10^4$ GeV$^2$ have this slope which confirms
the validity of Eq. (6).
 
Furthermore , substituting $\lambda = 0.42$ in Eq. (7),
one obtains $\langle Q^2 \rangle = 0.9 \; 10^4$ GeV$^2$ which agrees with
the asymptotic value reached at high energy
from explicit numerical integration of the diferential
cross section using the MRST98 partons (see Fig. 4a).

Assuming that the mean inelasticity $\langle y \rangle$ could
be experimentally determined,  
Eq. (6) can be inverted to obtain directly the $F_2$ slope
parameter $\lambda$ at small $x$ for $Q^2$ around $M_W^2$,
which is outside the kinematic limits of present accelerators.

We have to stress that the measurement of $\langle y \rangle$ will
not be easy to obtain because it would require the determination
of the final state energy corresponding to both, the nuclear cascade
and the lepton. If electron neutrinos are detected, 
the showers that come out of charged 
current interactions will have a different character depending on $y$  
because the neutrino induced electromagnetic (hadronic) shower arising 
at the leptonic (hadronic) vertex carries a fraction $1-y$ ($y$)
of the neutrino energy.
In particular it has been recently suggested that it may be 
possible to measure $y$ by using the radio technique \cite{avz99}. 
This tecnique relies of the detection of coherent radio pulses from the 
excess charge in the induced showers in a dense medium such as ice. 
As the radiation is coherent the angular structure of the distribution 
is sensitive to the longitudinal development of the shower. The 
sensitivity to $y$ arises from the elongation of the showers that 
develop at the electron vertex because of the 
Landau-Pomeran\v cuck-Migdal (LPM) \cite{LPM} effect \cite{avz99}. 

For the case of muon neutrinos in charged current interactions, 
both the muon and the nuclear shower in 
the final state would have to be detected.
As the muon energy loss is proportional to the muon energy above 1 TeV, 
this may not be out of question in a future large conventional neutrino
telescope such as IceCube \cite{ICE} although many difficulties can be
forseen.
In any case the ratios of rates detected with muons to those detected 
with showers should be dependent on $y$ for a given high energy neutrino flux.

\section{Conclusions}
\indent

The measurement of $\langle y \rangle$ at high energies in high energy
neutrino interactions would give direct
information on the behavior of the sea quarks distribution functions
in the nucleon in a kinematic range unreachable to the most powerful
accelerators, independently of the neutrino flux.
We have presented a simple analytical parametrization
which relates $\langle y \rangle$ with $\lambda$, the $x$ slope of
$F_2$ at low $x$ and high $Q^2$. 

We have also pointed out
that neutrino telescopes are sensitive to the value of
$\langle y \rangle$ and
the possibility
to measure $y$ from the interference pattern of the radio-signals
of the showers produced in the neutrino interaction.

Some other interesting information can be obtained from the measurement
of $y$. For example, it has been proposed that supersymmetric effects,
as $R$-parity violation could increase $\langle y \rangle$ with energy
significantly \cite{CARENA}. Also the effect of extra dimensions
in neutrino-nucleon interactions \cite{jain} should be manifested through
the modification of $\langle y \rangle$ \cite{Kachelriess}.

\vspace{1cm}
\hspace{1cm} \Large{} {\bf Acknowledgements}    \vspace{0.5cm}

\normalsize{}
This work was supported by Xunta de Galicia
under grant PGIDT00PXI20615PR and CICYT under grant AEN99-0589-C02-02).

\newpage

\vspace{0.5cm}

\hspace{1cm} {\Large{\bf Figure captions}}    \vspace{0.5cm}

{ \bf Figure 1.} The total neutrino-nucleon deep inelastic cross section
as a function of neutrino energy in lab frame for a) NC and CC neutrino
and antineutrino interactions with MRST98 partons and b) CC neutrino
interactions with MRST98 and GRV98 parton densities.  

\vspace{0.5cm}

{ \bf Figure 2.} The average inelasticity $y$
in deep inelastic neutrino-nucleon
interaction as a function of laboratory neutrino energy
for a) NC and CC neutrino
and antineutrino interactions with MRST98 partons and
b) CC neutrino interaction
with MRST98 and GRV98 parton densities.  

\vspace{0.5cm}

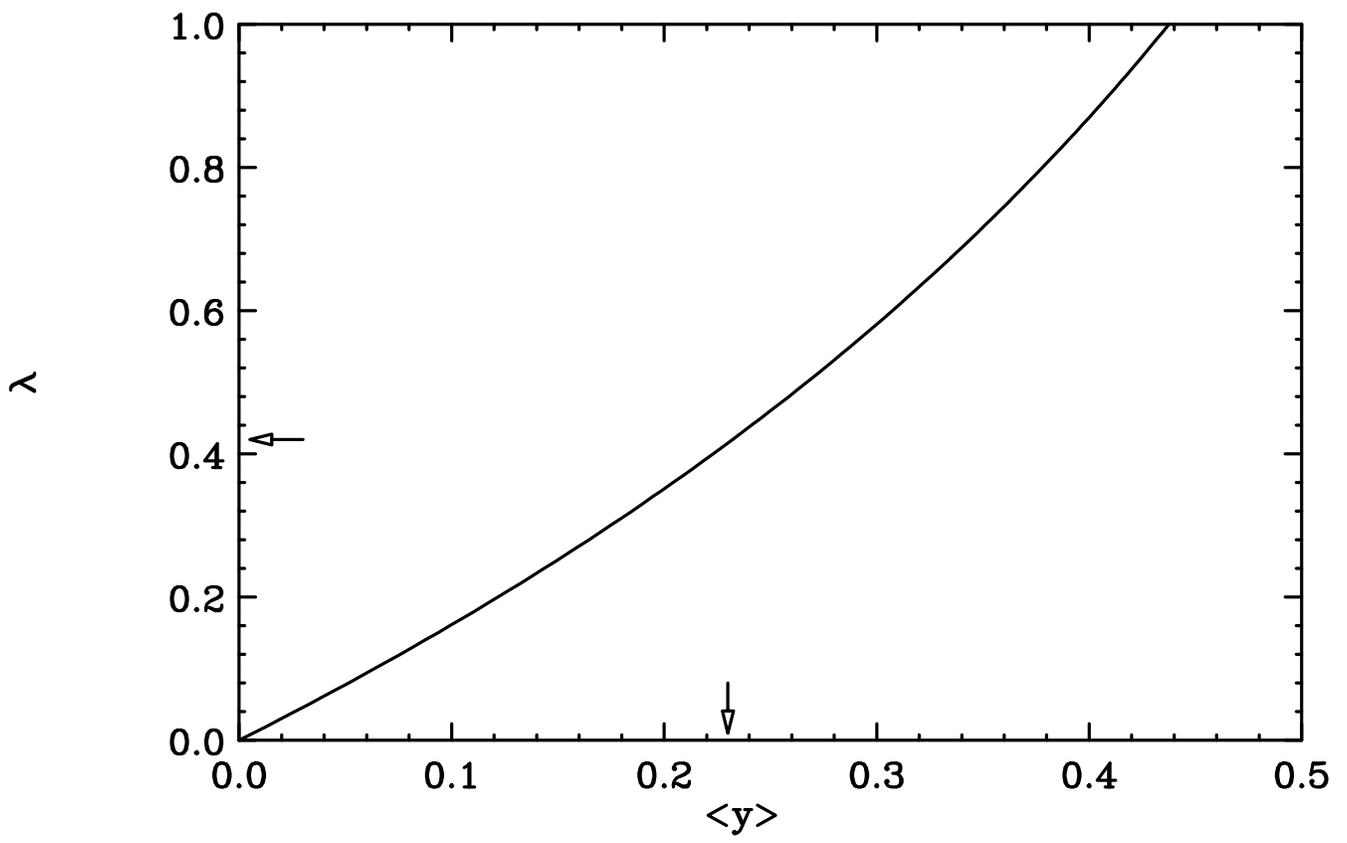
{ \bf Figure 3.} The parameter $\lambda$ ($F_2 \sim x^{-\lambda}$)
at low $x$ and high $Q^2 \simeq M_W^2$ as a function of the  
average inelasticity in CC $\nu N$ deep inelastic interaction.

\vspace{0.5cm}

{ \bf Figure 4.} a) The average $Q^2$ in deep inelastic
CC neutrino-nucleon interaction as a function of the laboratory
neutrino energy.
b) The average $x$ in deep inelastic
CC neutrino-nucleon interaction as a function of the laboratory
neutrino energy.

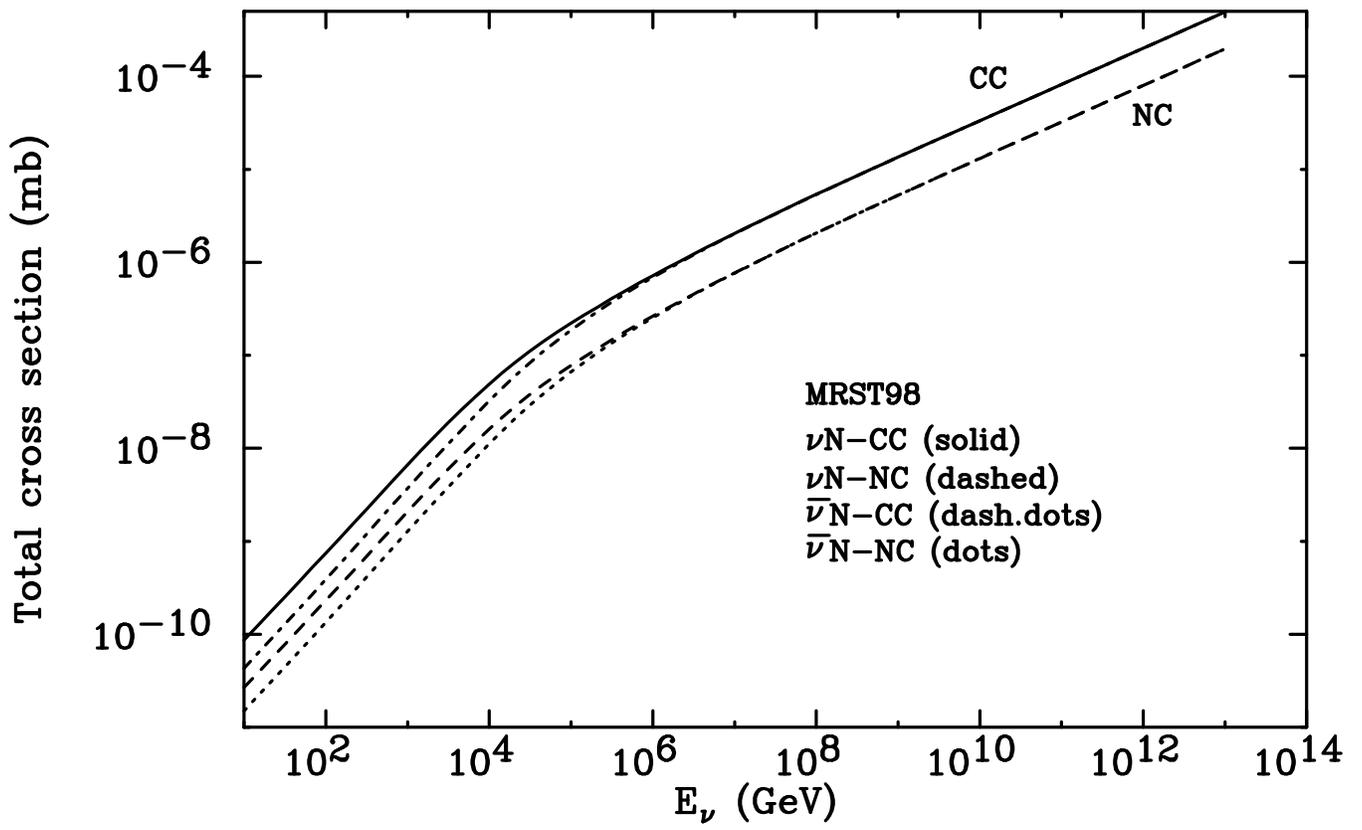
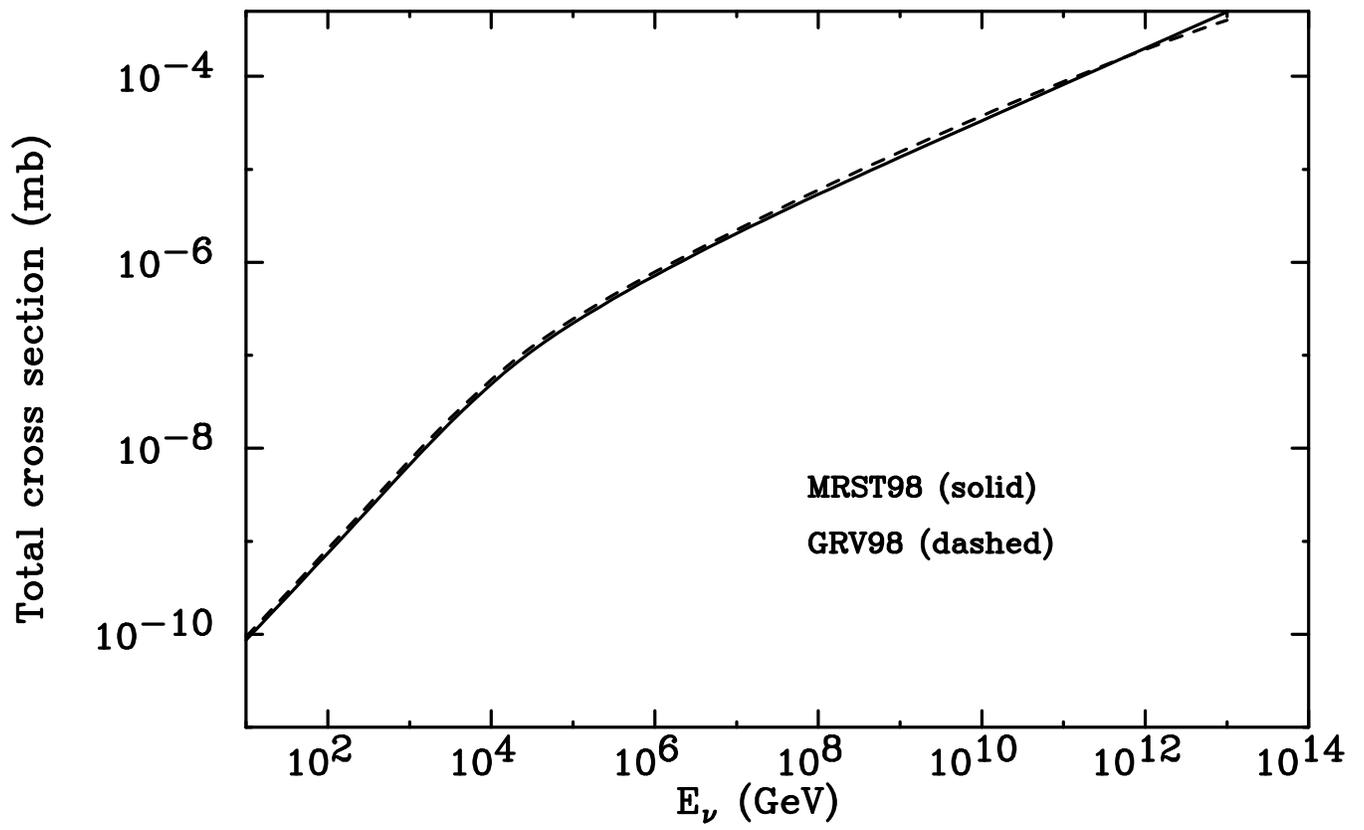
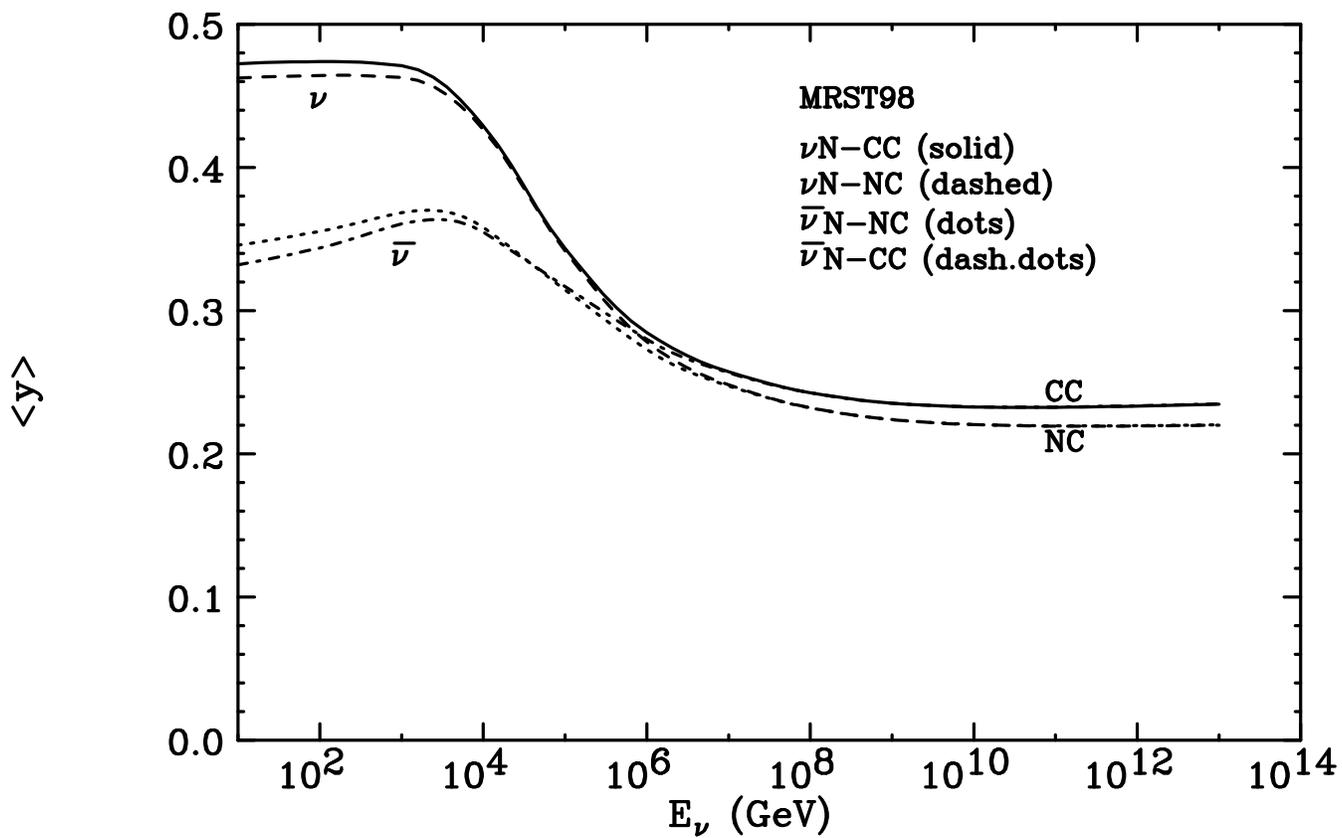
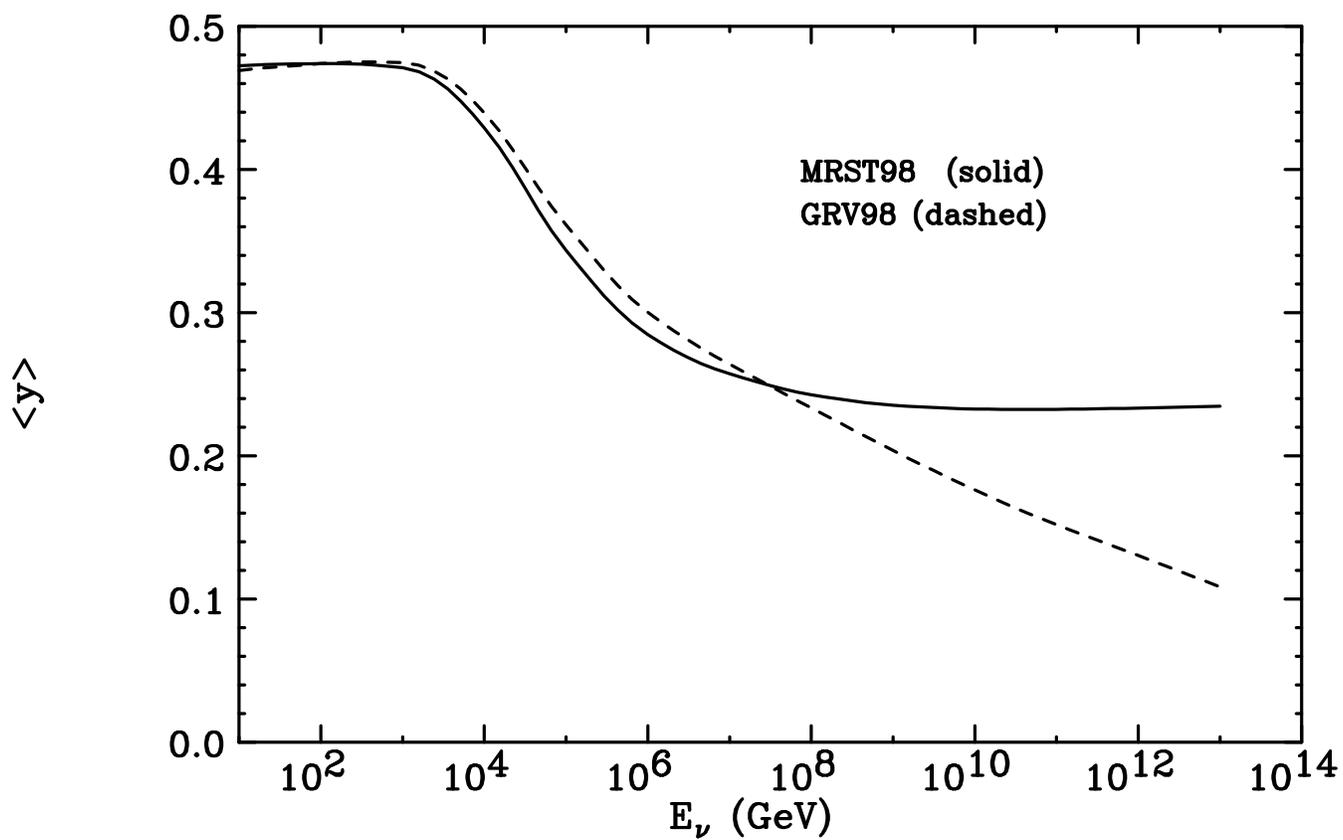
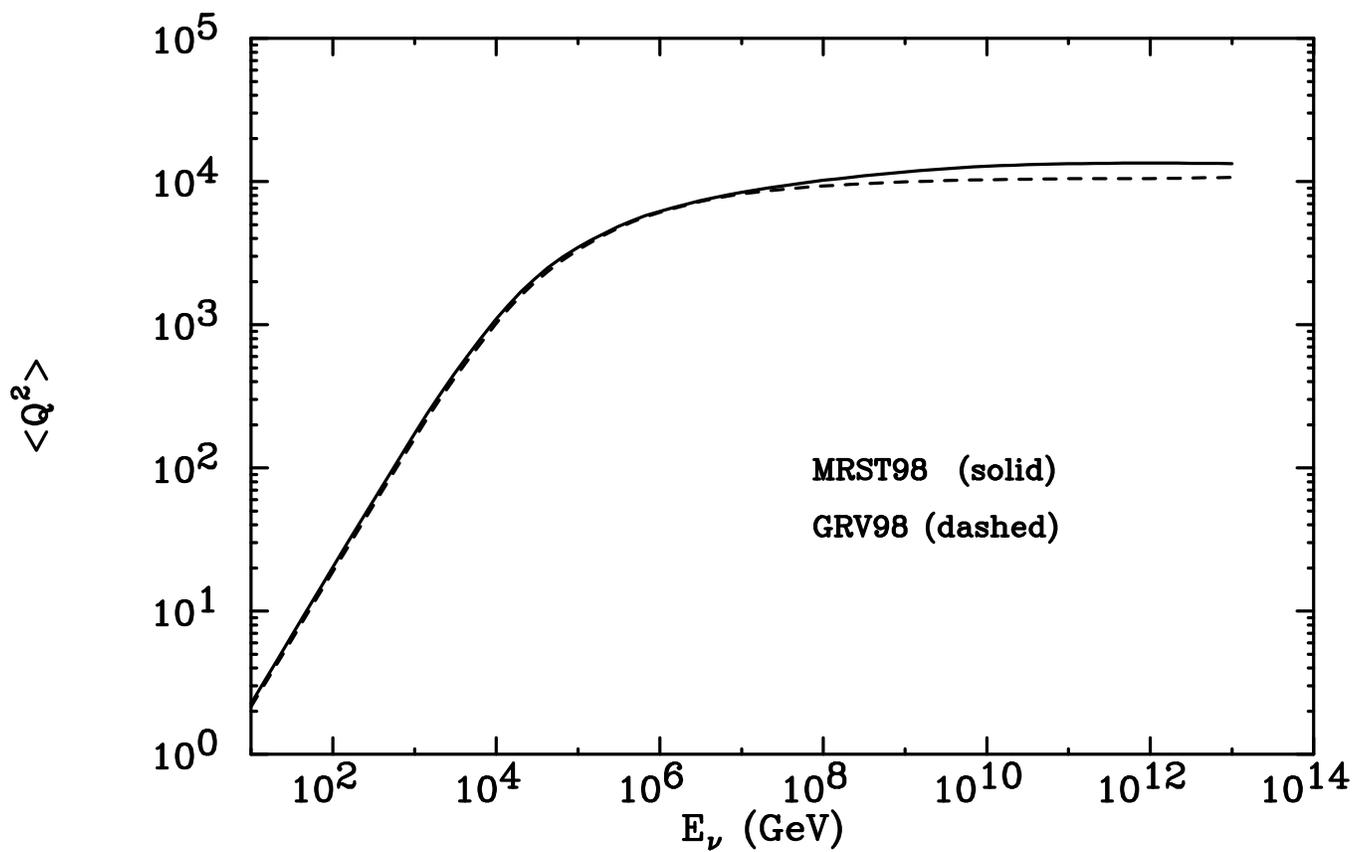
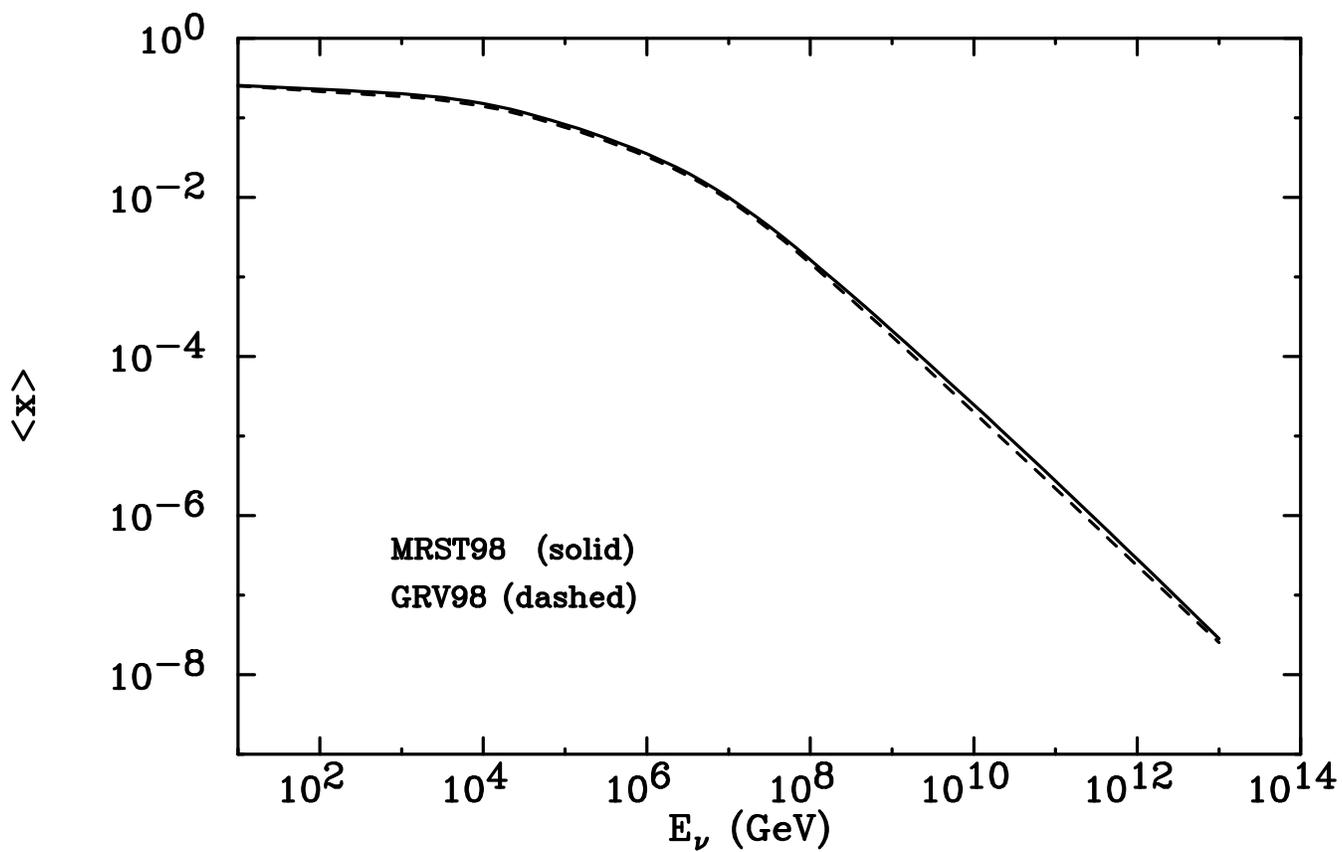
\end{document}